\documentclass{emulateapj}
\usepackage{./style_files/aas_macros}
\usepackage{natbib}
\usepackage{amsmath,amssymb,amsfonts}

\usepackage[plainpages=false, colorlinks=true, anchorcolor=blue, citecolor=blue, linkcolor=blue, bookmarks=false]{hyperref}
\usepackage{lineno}
\usepackage{graphicx}
\usepackage{color}
\usepackage{endnotes}
\usepackage{amssymb}
\usepackage{amsmath,bm}
\usepackage{times}
\usepackage{subfigure}
\usepackage{verbatim}
\usepackage{float}
\usepackage{cleveref}
\usepackage[normalem]{ulem}
\usepackage{xcolor}

\newcommand{\beq}{\begin{equation}}
\newcommand{\eeq}{\end{equation}}
\newcommand{\bdm}{\begin{displaymath}}
\newcommand{\edm}{\end{displaymath}}
\newcommand{\bea}{\begin{eqnarray}}
\newcommand{\eea}{\end{eqnarray}}

\newcommand{\cm}{\textcolor{black}}

\newcommand{\Msun}{M_\odot}
\def\lsim{\lower.5ex\hbox{$\; \buildrel < \over \sim \;$}}

\begin{document}

\title{The Local Nanohertz Gravitational-Wave Landscape From Supermassive Black Hole Binaries}

\author{Chiara M. F. Mingarelli\altaffilmark{1,2,3, $\dagger$}, T. Joseph W. Lazio\altaffilmark{3}, Alberto Sesana\altaffilmark{4}, Jenny E. Greene\altaffilmark{5}, Justin A. Ellis\altaffilmark{3}, Chung-Pei Ma\altaffilmark{6},  Steve Croft\altaffilmark{6}, Sarah Burke-Spolaor\altaffilmark{7,8}, Stephen R. Taylor\altaffilmark{3} }

\altaffiltext{$\dagger$}{mingarelli@mpifr-bonn.mpg.de}

\affil{$^{1}$Max Planck Institute for Radio Astronomy, Auf dem H\"{u}gel 69, D-53121 Bonn, Germany}
\affil{$^2$TAPIR, MC 350-17, California Institute of Technology, Pasadena, California 91125, USA }
\affil{$^3$Jet Propulsion Laboratory, California Institute of Technology, 4800 Oak Grove Drive, Pasadena, CA 91109, USA}
\affil{$^4$School of Physics and Astronomy and Institute of Gravitational Wave Astronomy, University of Birmingham, Edgbaston, Birmingham B15 2TT, United Kingdom}
\affil{$^5$Department of Astrophysical Sciences, Princeton University, Princeton, NJ 08544, USA}
\affil{$^6$Department of Astronomy, University of California, Berkeley, 501 Campbell Hall \#3411, Berkeley, CA 94720, USA}
\affil{$^7$Department of Physics and Astronomy, West Virginia University, Morgantown, West Virginia 26506, USA}
\affil{$^8$Center for Gravitational Waves and Cosmology, West Virginia University, Chestnut Ridge Research Building, Morgantown, WV 26505}

\begin{abstract}
Supermassive black hole binaries (SMBHBs) in the 10 million to 10 billion $\Msun$ range form in galaxy mergers, and live in galactic nuclei with large and poorly constrained concentrations of gas and stars. There are currently no observations of merging SMBHBs--- it is in fact possible that they stall at their final parsec of separation and never merge. While LIGO has detected high frequency GWs, SMBHBs emit GWs in the nanohertz to millihertz band. This is inaccessible to ground-based interferometers, but possible with Pulsar Timing Arrays (PTAs). Using data from local galaxies in the 2 Micron All-Sky Survey, together with galaxy merger rates from Illustris, we find that there are on average $91\pm7$ sources emitting GWs in the PTA band, and $7\pm2$ binaries which will never merge. Local unresolved SMBHBs can contribute to GW background anisotropy at a level of $\sim20\%$, and if the GW background can be successfully isolated, GWs from at least one local SMBHB can be detected in 10 years.
\end{abstract}

\maketitle

Supermassive black holes (SMBHs) are widely held to exist at the heart of massive galaxies \citep{Ferrarese:2005}. Galaxy mergers should form SMBH binary (SMBHB) systems, which eventually emit gravitational waves (GWs) and merge \citep{bbr80}. Galaxy mergers are a \cm{fundamental part} of hierarchical assembly scenarios, forming the backbone of current structure formation models. Thus, the detection of GWs from merging SMBHs would have fundamental and far-reaching importance in cosmology, galaxy evolution, and fundamental physics, not accessible by any other means. 

\cm{Pulsar Timing Arrays (PTAs) can detect nanohertz GWs by monitoring radio pulses between millisecond pulsars, which are highly stable clocks. GWs change the proper distance between the pulsars and the Earth, thus inducing a delay or advance of the pulse arrival times}. The difference between the expected and actual arrival times of the pulses -- the timing residuals -- carries information about the GWs that is extracted by cross-correlating the pulsar residuals \citep{ew75, s78, det79, hd83}. Current PTAs include European PTA \citep{ltm+15} (EPTA), the North American Nanohertz Observatory for Gravitational Waves (NANOGrav) \citep{ArzoumanianEtAl:2016} the Parkes PTA \citep{ShannonEtAl:2015}, and the International PTA (IPTA) \citep{VerbiestEtAl:2016}, the latter being the union of the former three.

\cm{Here we introduce a bottom-up approach to constructing both realistic GW skies and future IPTA projections: we use IPTA pulsars with their real noise properties, and galaxies from the 2 Micron All Sky Survey (2MASS) \citep{2mass}, together with galaxy merger rates from Illustris \citep{RodriguezGomez+:2015, Illustris2014}, to form multiple probabilistic realizations of the local GW Universe.}
In each realization, we search for SMBHB systems which emit continuous GWs (CGWs) in the PTA band, and also compute their contribution to the nanohertz GW background (GWB) and its anisotropy \citep{msmv13, SchutzMa:2016}. We report on the physical properties of the most frequently selected SMBHBs and their host galaxies, and estimate their time to detection.
 
\section*{Galaxy Selection}
SMBHB merger timescales can be of the order $10^9$~yrs after the galaxy merger, and therefore morphological merger signatures can be difficult to identify. 
Our focus here is on massive early-type galaxies, as these are likely to have formed from major mergers and would therefore host SMBHBs with \cm{approximate} mass ratios \cm{of} $0.25\leq q \leq 1$. 

To approximate a mass selection, we select in the $K-$band using the 2MASS \citep{2mass} Extended Source Catalog (XSC) \citep{JarrettEtAl:2000}, following the procedure outlined in detail in Ref \citep{MaEtAl:2014}, but to a distance of 225\,Mpc and over the full sky. We do not excise the galactic plane, but
there are fewer reliable sources there. To approximate a
mass selection, we select in the $K-$band using the 2MASS \citep{2mass}
 Extended Source Catalog (XSC) \citep{JarrettEtAl:2000}.
We convert from the 2MASS K-band luminosity to the stellar mass via $\log_{10}(M_*) = 10.58-0.44\,(M_K+23)$, 
appropriate for early-type galaxies \citep{Cappellari:2013}, and apply a K-band cut $M_K\leq-25$ to select galaxies with stellar mass of $M_*\gtrsim 10^{11}~\Msun$, since these are likely to host SMBHBs \citep{RosadoSesana:2014}. \cm{At distances $> 225$\,Mpc, 2MASS itself begins to become incomplete at $M_K \sim -25$. Although our sample is not formally volume-complete within our chosen mass and distance cuts, it includes the majority of massive, nearby galaxies, which are expected to dominate the signal for PTAs.}

\cm{This process creates a galaxy catalog with 5,110 early-type galaxies. In Ref \citep{MaEtAl:2014, SchutzMa:2016}, we find that 33 of these galaxies contain dynamically measured SMBHs, Figure \ref{fig:2mass}. Moreover, we manually add a further 9 galaxies from 2MASS: NGC~4889, NGC~4486a, NGC~1277, NGC~1332, NGC~3115, NGC~1550, NGC~1600, NGC~7436 and A1836~BCG, which did not make the luminosity cut, but also host dynamically measured SMBHs.}

\cm{The SMBHB total mass, $M=M_1+M_2$, is estimated by taking the stellar mass to be the bulge mass, and applying the $M_\bullet-M_{\mathrm{bulge}}$ empirical scaling relation from Ref \citep{mm13}, with scatter $\epsilon_0$, described in Methods. Dynamically measured SMBH masses fixed.}

\section*{Probability of a Galaxy Hosting a SMBHB in the PTA Band} 

\cm{We attack this problem by estimating two quantities: the probability that a SMBHB is in the PTA band, and the probability that a given galaxy has a SMBHB. The former is estimated via quantities which are derived from 2MASS, e.g. each galaxy's stellar mass, $M_*$, and through $M_\bullet-M_\mathrm{bulge}$, the SMBHB mass M. The latter is obtained from the cumulative galaxy-galaxy merger rates from the Illustris cosmological simulation project \citep{RodriguezGomez+:2015, Illustris2014}.}

\cm{A realization of a the local nanohertz GW sky is created as follows. The first quantity is the probability that a SMBHB is in the PTA band, $t_c/T_z$, where $t_c = 5/256 (\pi f)^{-8/3} \mathcal{M}_c^{-5/3}$ is the time to coalescence of the binary and $\mathcal{M}_c^{5/3} = \left[ q/(1+q)^2 \right] M^{5/3}$ is the chirp mass. }

\cm{In each realization, all galaxies in the sample are assigned a SMBHB with total mass $M$, and mass ratio $q$ drawn from a log-normal distribution in $[0.25,1]$. The time to coalescence is computed from $1$~nHz, which is the beginning of the PTA band. $T_z$ is the effective lifetime of the binary: this is the sum of the dynamical friction \citep{BinneyTremaine} ($t_\mathrm{df}$) and stellar hardening \citep{sk15} ($t_\mathrm{sh}$) timescales (details in Methods).}
 
\cm{The second quantity is the probability that a galaxy hosts SMBHB. This is computed via the Illustris cumulative galaxy-galaxy merger rate, taken at the beginning of the binary evolution. This redshift $z$ is calculated at $T_z$ with Planck cosmological parameters \citep{Planck2015}. The merger rate is then multiplied by time elapsed since then, $T_z$. Thus, probability of galaxy $i$ hosting a SMBHB in the PTA band, $p_i$}
\beq
\label{eq:dNdT}
p_i= \frac{t_{c,i}}{T_z}\int_{0.25}^{1} \mathrm{d}\mu_* \frac{dN}{dt}(M_*, z, \mu_*) T_z\, .
\eeq
\cm{Calculating $T_z$ is essentially rewinding the SMBHB evolution: starting in the GW emission phase, we compute how long the binary spent in a stellar hardening phase, and then in a dynamical friction phase, for binary separations out to the effective radius of the galaxy. We can only rewind 12.5 Gyrs $(z=4)$, as this is the maximum $z$ from Ref \citep{RodriguezGomez+:2015}. If $T_z>12.5$~Gyr (equivalently $z>4$), binary has likely stalled, and we set $p_i=0$.}

The total number of SMBHBs emitting GWs in the PTA band, for each Monte Carlo realization, is the sum of all these probabilities: $N_{\rm SMBHB} = \sum_i p_i$. This number varies from realization to realization. We draw $N_{\rm SMBHB}$ galaxies from the galaxy catalog according to the probability distribution, ${\cal P}=p_i/N_{\rm SMBHB}$.

\cm{For each of these selected galaxies, we compute the inclination and polarization-averaged strain,
\beq 
\label{eq:h}
h=\sqrt{\frac{32}{5}}\frac{\mathcal{M}_c^{5/3}}{D_L} \left[\pi f(1+z)\right]^{2/3}\, ,
\eeq
assuming circular binary orbits, and $f$ is the GW frequency. Each of the probabilistically selected galaxies hosts a SMBHB in the PTA band with $f>1$~nHz, where the evolution is assumed to be dominated by GW emission. We assign each SMBHB a GW frequency \citep{p64}, $f$:
\beq
\label{eq:freq}
f=\pi^{-1}\mathcal{M}_c^{-5/8}\left[ \frac{256}{5}(t_c-t)\right]^{-3/8} \, ,
\eeq
by drawing $t_c$ from a \cm{uniform distribution in $[26~\rm{Myr}, 100~\rm{yr}]$, which is the time to coalescence of a SMBHB with $M_1=M_2=10^9~\Msun$} from 1 nHz and 100 nHz, respectively\footnote{Formally we set $t_c=0$ and sample in $-t$}. Galaxy distances are estimated via techniques outlined in the Methods section}. \cm{We use the approximation $z = 0$ only in Equation \eqref{eq:h}, since GW sources are all $<225$~Mpc.}

\section*{Projections for the IPTA}
\begin{figure*}
\begin{center}
       \subfigure[Distribution of galaxies in our catalog]
  	{%
	\includegraphics[width=0.45\textwidth]{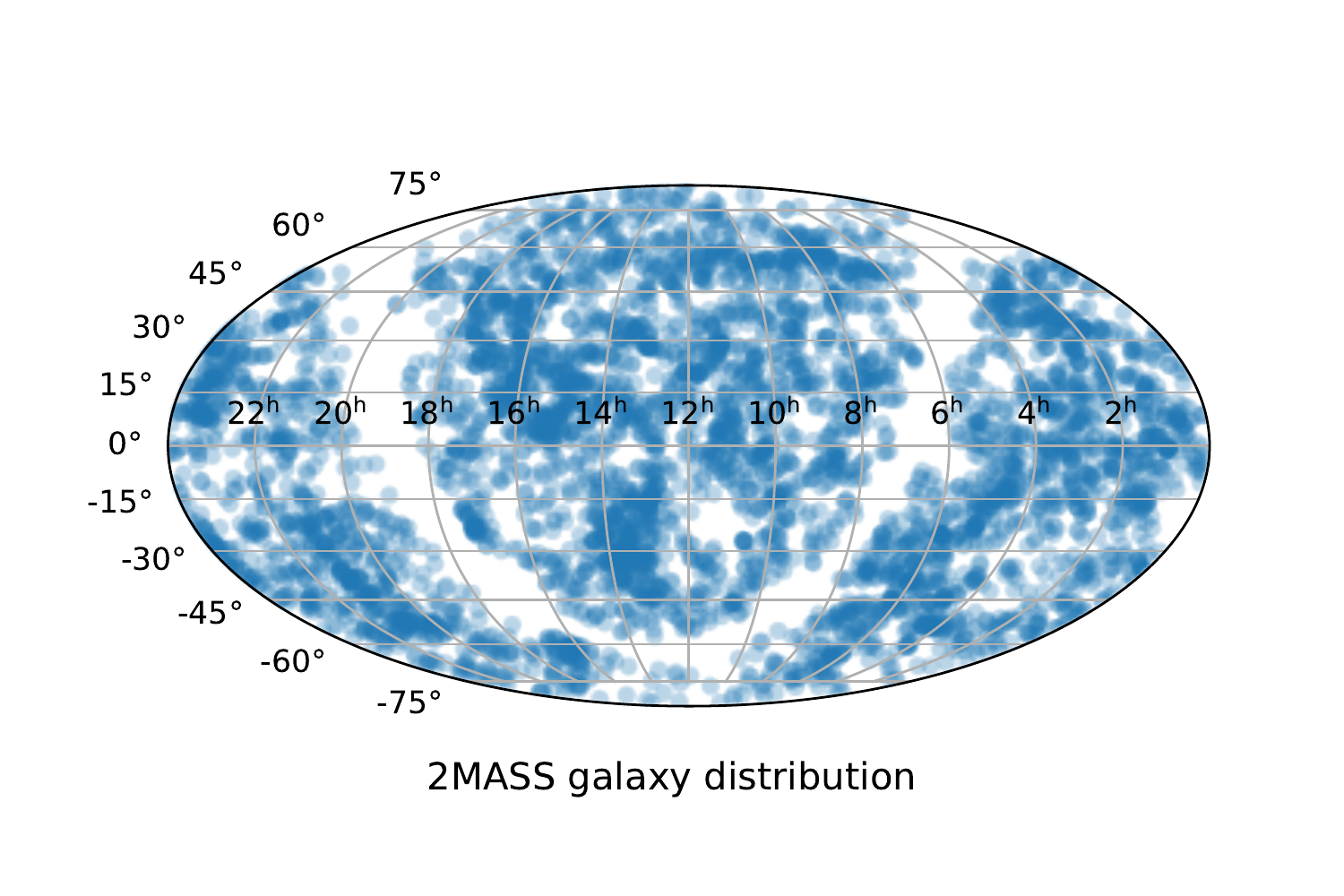}
        \label{fig:2mass}
        }    
   \subfigure[All-sky GW strain sensitivity map at $f=3.79$ nHz]
  	{%
	\label{fig:allSky}
	\includegraphics[width=0.45\textwidth]{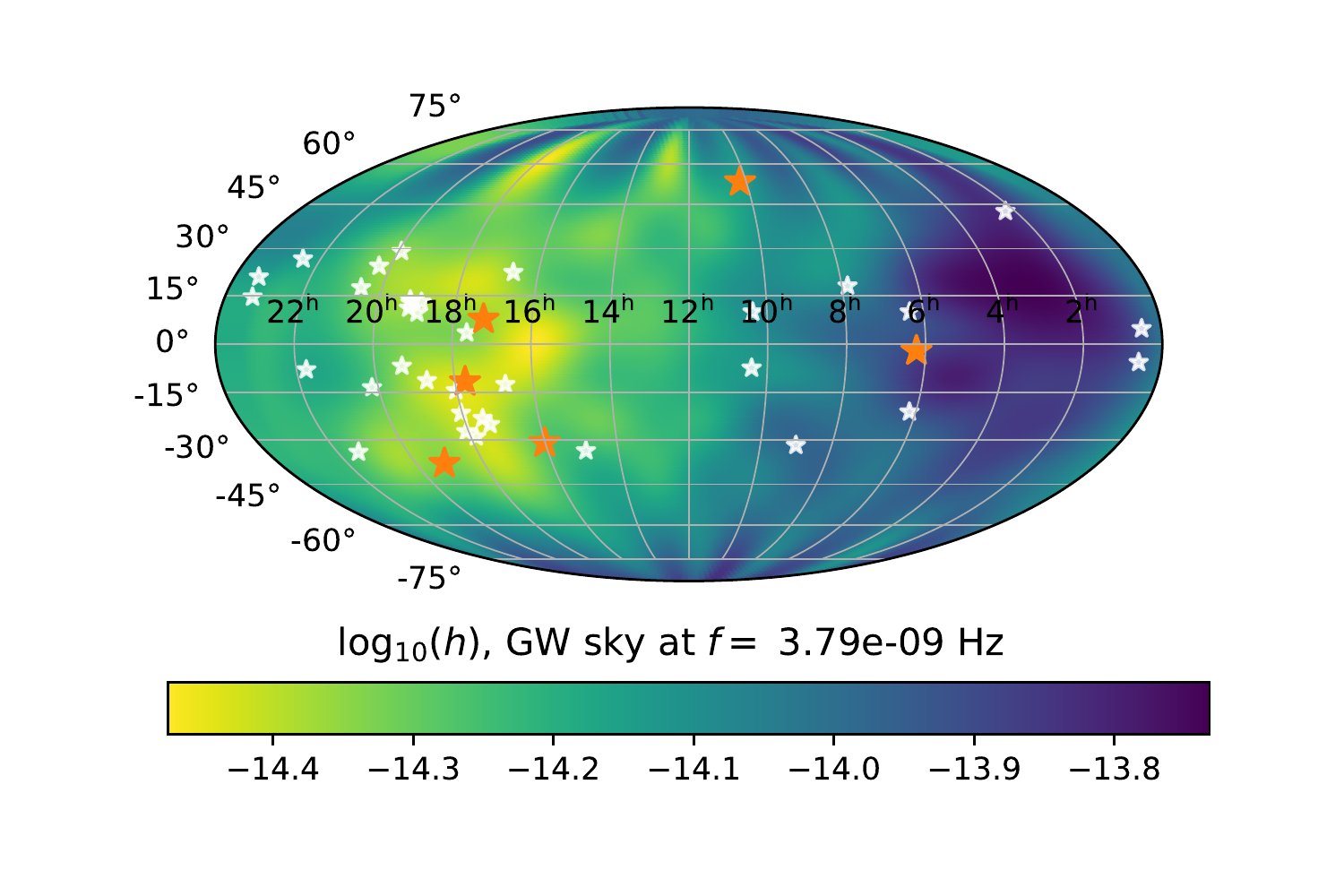}
        }

      \subfigure[All detected SMBHB host galaxies]
  	{%
	\label{fig:detection}
	\includegraphics[width=0.45\textwidth]{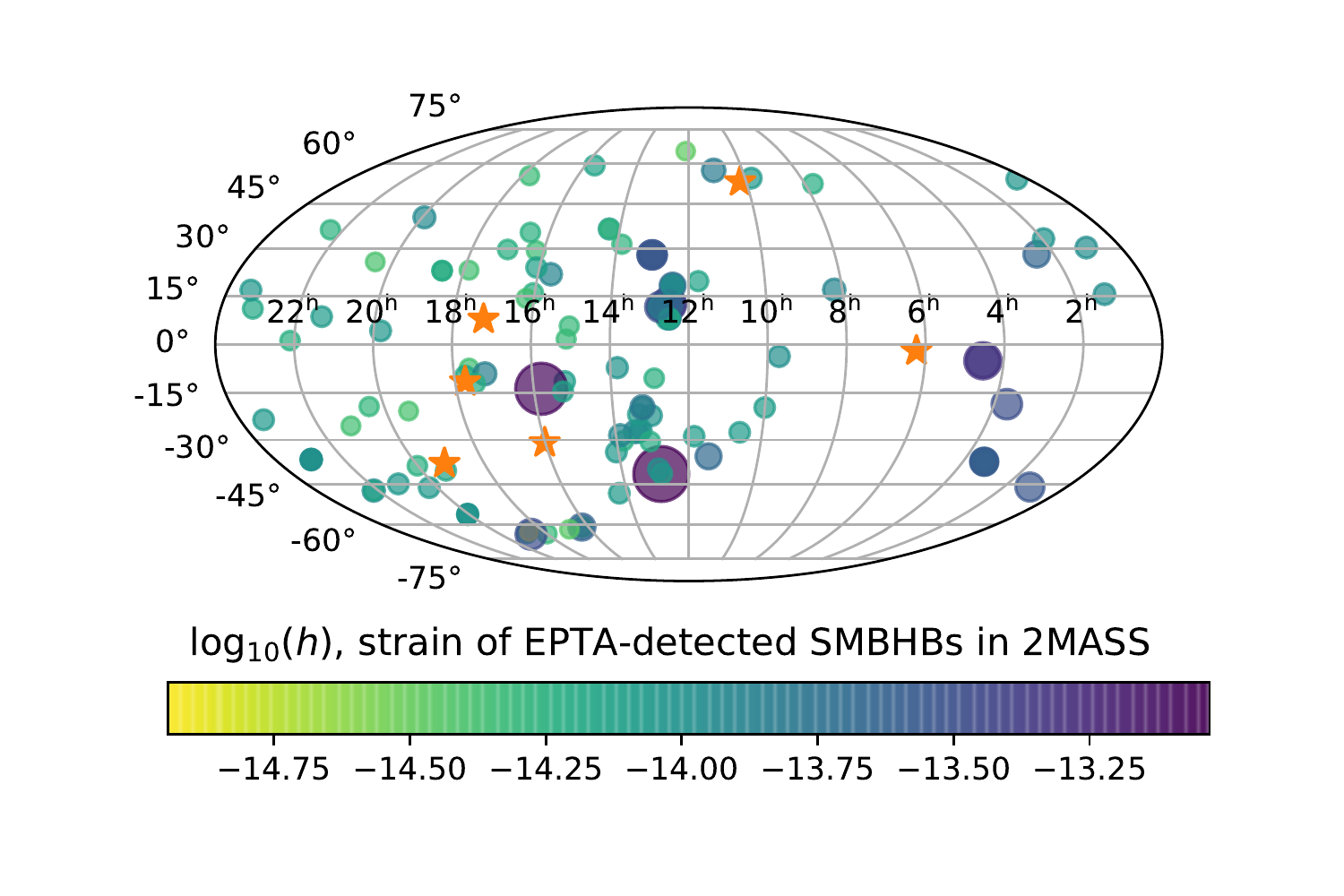}
        }    
    \subfigure[Sky-averaged GW strain]
  	{%
	\label{fig:skyAvg}
	\includegraphics[width=0.45\textwidth]{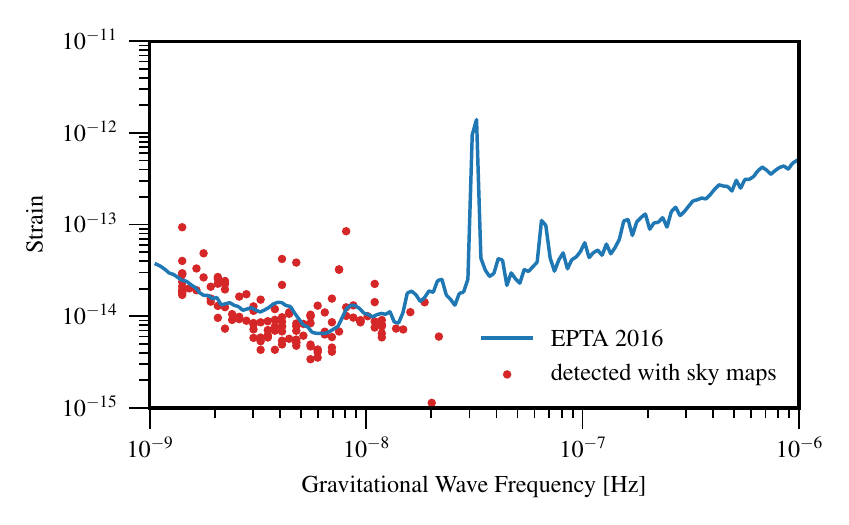}
        }      

  \end{center}
\caption{{\bf Importance of pulsar positions for gravitational wave detection} (a) All selected 2MASS galaxies. (b) Example of an all-sky gravitational-wave (GW) strain map using our own reprocessing of data from Ref \citep{bps+15}. The strain upper limit on the sky behind the best 6 pulsars (orange stars) is four times more constraining than the 35 other pulsars (white stars).  (c) We find 131 SMBH binaries emitting continuous GWs in the PTA band across all GW frequencies, using all-sky GW sensitivity maps, as in (b). The size and color of the circle indicates the relative strain of the source. (d) The sky-averaged strain upper limit of continuous GW sources, Ref \citep{bps+15}. \cm{Only 34 of the 131 detected sources (red dots) lie above the upper limit curve, demonstrating that sky-averaged strain underestimates the number of detectible GW sources by a factor of $\sim 4$. For IPTA projections, see Table \ref{tab:ipta}.}}
\label{fig:allSources}
\end{figure*}

Continuous nanohertz GW upper limits come in two flavors: 1) as a function of GW source sky location, GW frequency, and pulsar sky location (Figure \ref{fig:allSky}), and 2) a sky-averaged strain upper limit as a function of GW frequency, which averages over all pulsar sky locations, Figure \ref{fig:skyAvg}. Since upper limit curves, and not detection curves, were reported by Ref \citep{bps+15}, we use the upper limit as a proxy for detection to underline the importance of the pulsar and SMBHB sky location. Indeed, the PTA response to CGWs is maximal when the GW source lies very close to the pulsar \citep{det79, ms14}.

For IPTA projections, we are more rigorous, constructing an IPTA-like array to estimate the time to detection of CGW sources in 2MASS. We start with the 47 IPTA pulsars \citep{VerbiestEtAl:2016}, and from 2016 onward we add four pulsars per year from sky locations accessible with IPTA telescopes, using the median white noise uncertainty of 300 ns. 

Detection probability curves are computed using the $\mathcal{F}_{p}$ statistic \citep{esc12} and False Alarm Probabilities \citep{abb+14} (FAPs) of $5\times 10^{-2}$, $3\times 10^{-3}$, and $1\times10^{-4}$. For Gaussian distributions, these 1-FAPs are $2\sigma$, $3\sigma$ and $3.9\sigma$ respectively.

\section*{Gravitational Wave Backgrounds}
The search for the GWB is ongoing \citep{ArzoumanianEtAl:2015, desvignes+:2016, VerbiestEtAl:2016, ArzoumanianEtAl:2016, ltm+15, ShannonEtAl:2015} with detection expected in 5 to 7 years \citep{tve+16}. While most PTAs publish limits on isotropic GWBs, searching for and characterizing GWB anisotropy is gaining momentum \citep{TaylorEtAl:2015}. This is done by decomposing the sky on a basis of spherical harmonics \citep{msmv13,tg13, grt+14}, $Y_{\ell m} (\hat\Omega)$, and with pixel-bases \citep{cvh14}. 

In an effort to understand the contribution of these CGW sources to the GWB, we transform a GW sky realization to a GWB. This is done by casting each individual binary's strain contribution to the closest pixel in a HEALPix sky map, and computing the characteristic strain $h^2_c=\sum_k h^2_k f_k/\Delta f$, where $h_k$ is the strain of source $k$, $f$ is its GW frequency, and $\Delta f$ is the inverse observation time \citep{bps+15}. 

\cm{The total power on the sky is $4\pi$, and is decomposed as $P(\hat\Omega)=\sum_{\ell,m}c_{\ell m}Y_{\ell m} (\hat\Omega)$, where $\hat\Omega$ is the direction of GW propagation. Anisotropy in described terms of $ C_\ell = \sum_{m=-\ell}^{+\ell} |c_{\ell m}|^2/(2\ell+1) $, normalized to the isotropic component, $C_0$.  We calculate the angular power spectrum of a GW sky and take the monopole value as the contribution to the overall isotropic GWB.}

\begin{table*}
  \centering
  \begin{tabular}{c|c|c|c||c|c|c}

		&	\multicolumn{3}{c}{Sky-Averaged Strain}	& \multicolumn{3}{c}{All-sky GW strain sensitivity map } \\
	FAP	&	15 yrs	&	20 yrs	&	25 yrs  & 15 yrs	&	20 yrs	&	25 yrs \\
\hline
$5\times 10^{-2}$ 	& 2\% (0.1\%)	& 24\% (0.3\%)	& 100\% (0.8\%) & 8\% (0.4\%)	& 96\% (1\%)	& 100\% (3\%) \\

$3\times 10^{-3}$ 	& 0.5\% (0.03\%)	&   9\% (0.2\%)	 &  48\% (0.3\%) & 2\% (0.1\%)	&   36\% (0.8\%)	 &  100\% (1\%) \\
$1\times10^{-4}$	& 	0.3\% (0.01\%)	 &  4\% (0.08\%)	&   27\%(0.2\%)& 	1\% (0.04\%)	 &  16\% (0.3\%)	&   100\% (0.8\%)\\
\end{tabular}
\caption{{\bf Probability of detecting nearby continuous gravitational-waves with the International Pulsar Timing Array} The predictions are reported for different False Alarm Probabilities (FAP), total length of IPTA dataset, and if the pulsars are dominated by white (red) noise. This probability is the total number of sources from all realizations lying above a given detection curve, divided by the total number of realizations, with a maximum of 100\% (even though multiple sources may be detected; Figure \ref{fig:iptaProj}). For Gaussian distributions, 1-FAPs are $2\sigma$, $3\sigma$ and $3.9\sigma$. PTA datasets are now between 10 and 15 years long, \cm{hence 25 year projections are likely 10 years from now. Using an all-sky GW strain sensitivity map may add an additional factor of 4 in GW strain sensitivity, as seen in Figure \ref{fig:allSources}, which highlights the importance of pulsar positions on the sky with respect to the GW source.}}
\label{tab:ipta}
\end{table*}

\section*{Results}

\textbf{Galaxies hosting SMBHBs}
We compute the probability of each galaxy in the catalog containing a SMBHB emitting GWs in the PTA band, $f\geq 1$~nHz. We carry out multiple realizations of the galaxy catalog, sampling over BH mass, mass ratio, and time to coalescence. We find that on average, there are $91\pm7$ galaxies hosting SMBHBs in the PTA band and $7\pm2$ stalled SMBHBs, despite the inclusion of a stellar hardening phase \citep{q96} to overcome the final parsec problem (Supplementary Figure \ref{fig:GalStall}). 

Over multiple realizations, fewer than $1\%$ of GW skies hosted a currently detectable SMBHB. \cm{We also note that for 25 year datasets, binaries with $\mathcal{M}_c\sim10^9~\Msun$ are the likeliest to be detected, with more massive binaries being disfavored, Supplementary Figure \ref{fig:chirp}.}

Six pulsars dominate the EPTA's sensitivity to CGWs, leading to an upper-limit map of the sky that is approximately dipolar, with a factor of 4 more in strain sensitivity behind the best pulsars, Figure \ref{fig:allSources}. While we use the EPTA as an example, results are similar for NANOGrav, PPTA and the IPTA.

\begin{figure*}
\begin{center}
   
    \subfigure[Example realization where NGC 4472 was detected]
  	{
	\includegraphics[width=0.45\textwidth]{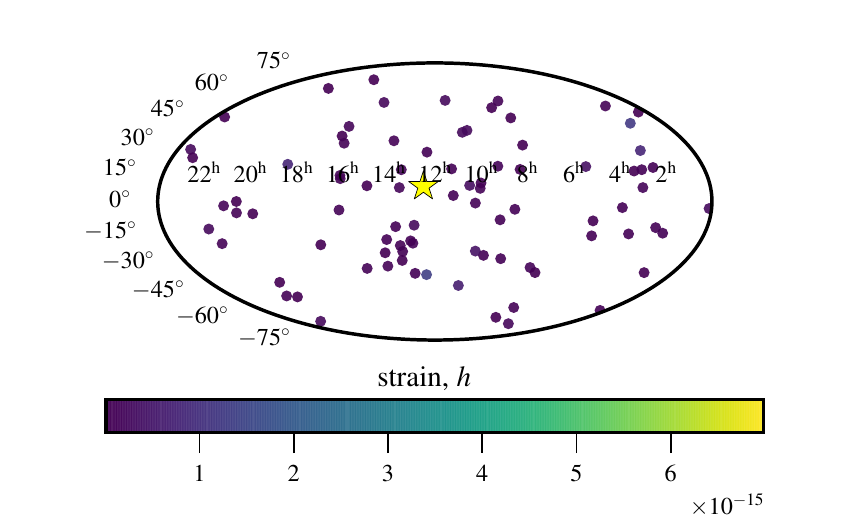}
	\label{fig:detected_src}
        }    
       \subfigure[This GW sky turned into a GW background]
  	{%
	\includegraphics[width=0.41\textwidth]{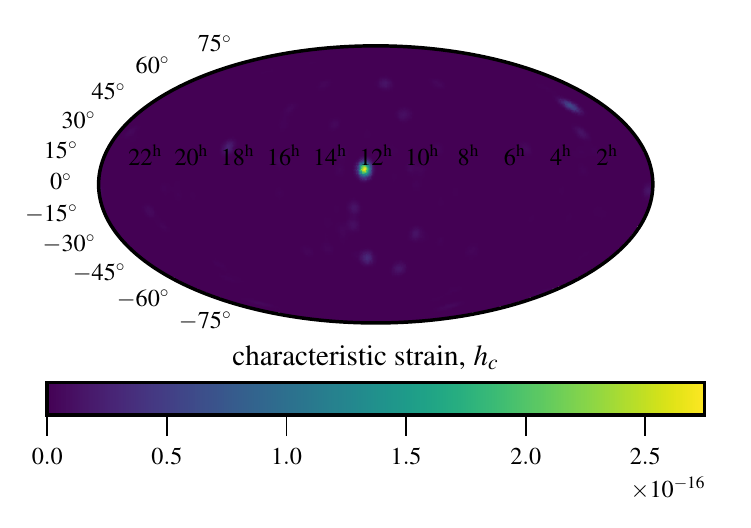}
	\label{fig:detected_backg}
        }

      \subfigure[GW background without NGC 4472]
  	{%
	\includegraphics[width=0.45\textwidth]{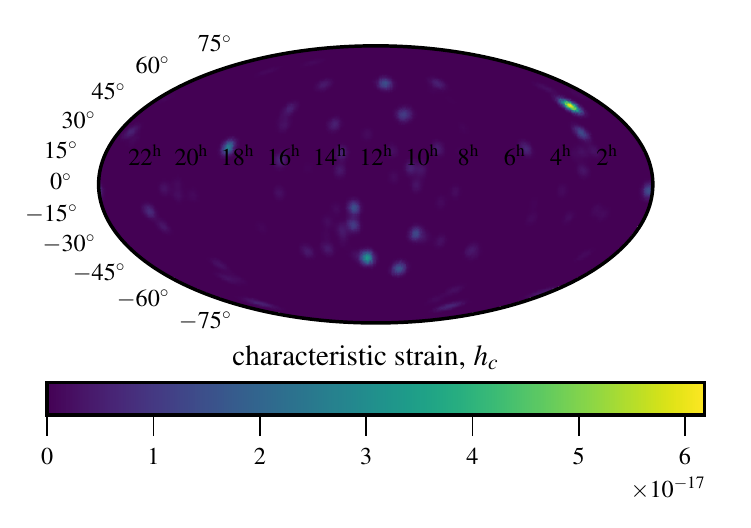}
	\label{fig:detected_Nosrc}
        }    
   \subfigure[Angular power spectrum of GW background]
  	{%
	\includegraphics[width=0.45\textwidth]{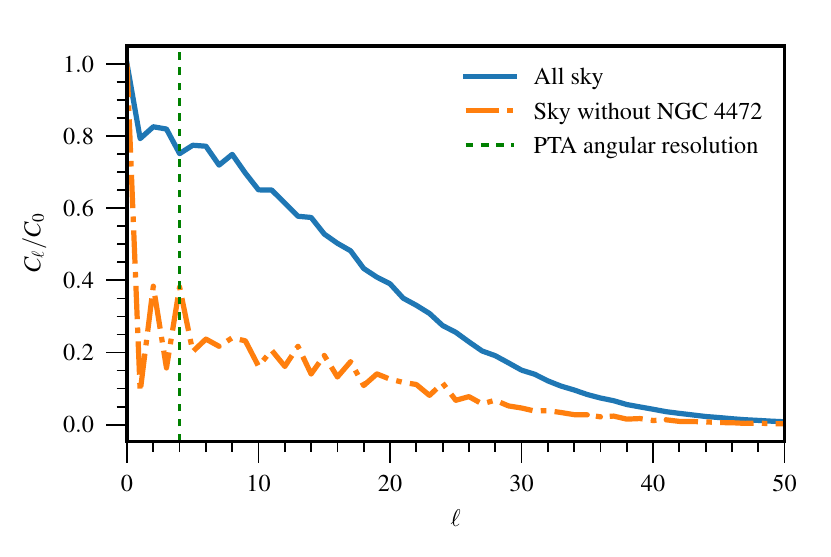}
	\label{fig:anis}
        }    
  \end{center}
\caption{{\bf The gravitational wave background from nearby sources} (a) NGC 4472 (yellow star) hosted a detectable SMBHB in one of the realizations. There were 86 other galaxies hosting SMBHBs with $f>1$~nHz, but these were too faint to be detected. (b) The 87 CGW sources from (a) turned into a GW background, assuming a 25-year PTA dataset. This background is dominated by NGC 4472. (c) The GW background without NGC 4472. (d) The angular power spectrum of the GW backgrounds both with and without NGC 4472, assuming a 25 year dataset. Even without the strong CGW source in NGC 4472, the other 86 SMBHBs produce anisotropy at the level $\ge $20\% up to $\ell=15$.}
\label{fig:allanis}
\end{figure*}

\textbf{Anisotropy and contributions to the GW background} 
\cm{In Figure \ref{fig:detected_src}, we show an example realization of the local nanohertz GW sky with a loud SMBHB, chosen at random. In this realization, NGC 4472 hosted a PTA-detectable SMBHB. Of course, NGC 4472 was only one of 87 galaxies in this realization, but it was the only one which contained a binary that was sufficiently loud to be detected.
Assuming an isotropic GWB with an amplitude of a few \cm{times} $10^{-16}$ and a 25 year dataset, we find that such a single CGW source contributes less than 1\% to the overall strain budget. This is to be expected, since a CGW is the ultimate anisotropy, and therefore contributes very little to an isotropic GWB. The strong CGW source does, however, dominate the angular power spectrum of the sky, Figure \ref{fig:anis}, where we measure GWB anisotropy. When the strong CGW source is removed, Figure \ref{fig:detected_Nosrc}, we find anisotropy from the superposition of GWs from undetected CGW sources at the level of $\sim 20\%$, dominating the angular power spectrum of the GWB up to $\ell \sim 10$, Figure \ref{fig:anis}.}

\textbf{Time to Detection} Evidence for nanohertz GWs will increase slowly and continuously. We estimate the time to detection of nearby GW sources with the IPTA, requiring a 95\% detection probability under different false alarm probabilities (FAP): $5\times 10^{-2}$, $3\times 10^{-3}$, $1\times10^{-4}$ ($2\sigma$, $3\sigma$ and $3.9\sigma$ for Gaussian distributions). This is done for 15, 20, and 25-year datasets, noting that current datasets are 10-15 years. Hence, 25-year predictions are for 10 years from now. Results for detections with $3\times 10^{-3}$ FAP are in Figure \ref{fig:iptaProj}, and summarized in Table \ref{tab:ipta}. 

We find that strong red noise in pulsar residuals greatly diminishes the chance of detecting CGW sources in the next 10 years. However, if the pulsar noise is white or if CGW sources can be extracted from an unresolvable GWB, then there is a 50\% chance of detecting a local CGW source with a $3\times 10^{-3}$ FAP in the next decade. Even more encouragingly, when one uses the all-sky GW strain sensitivity map for detections, a detection with $10^{-4}$ FAP is possible in 10 years, Table \ref{tab:ipta}.

\section*{Discussion}
Using the sky location and noise properties of IPTA pulsars, massive galaxies in 2MASS and galaxy merger rates from Illustris, we estimate when and where PTAs are likely to detect CGW sources in the local Universe. 
\cm{Over multiple realizations of the local GW sky, we find that $\ll1\%$ of GW sources would have been detected with current PTA data \citep{desvignes+:2016}, Figure \ref{fig:detection}, supporting the conclusions that current non-detection is unsurprising, Ref \citep{bps+15} .}

\cm{In making IPTA predictions, we did not include new telescopes which will come on line in the next 10 years, such as MeerKAT \citep{MeerKAT} and possibly SKA Phase 1 \citep{JanssenEtAl:2015}. }Both telescopes will greatly increase PTA sensitivity in the Southern hemisphere. 

Importantly, future IPTA detections depend on how successfully the GWB can be subtracted. The red noise in the pulsars is meant to emulate an unresolved GWB with $A=4\times 10^{-16}$, consistent with Ref \citep{SesanaEtAl:2016}. While some pulsars exhibit intrinsic red noise, the highest precision timers -- which contribute most to CGW sensitivity -- are broadly consistent with being white noise dominated. 

\begin{figure*}
\begin{center}
	\subfigure[Top SMBHB host galaxies (pessimistic) ]{
	\includegraphics[width=0.45\textwidth]{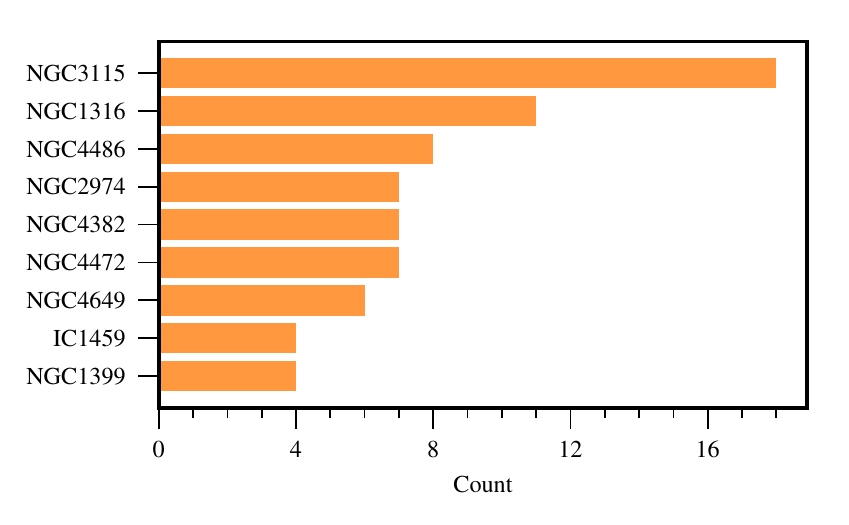}
	\label{fig:hitlist_red}
	}
	\subfigure[Top SMBHB host galaxies (optimistic)]{
	\includegraphics[width=0.45\textwidth]{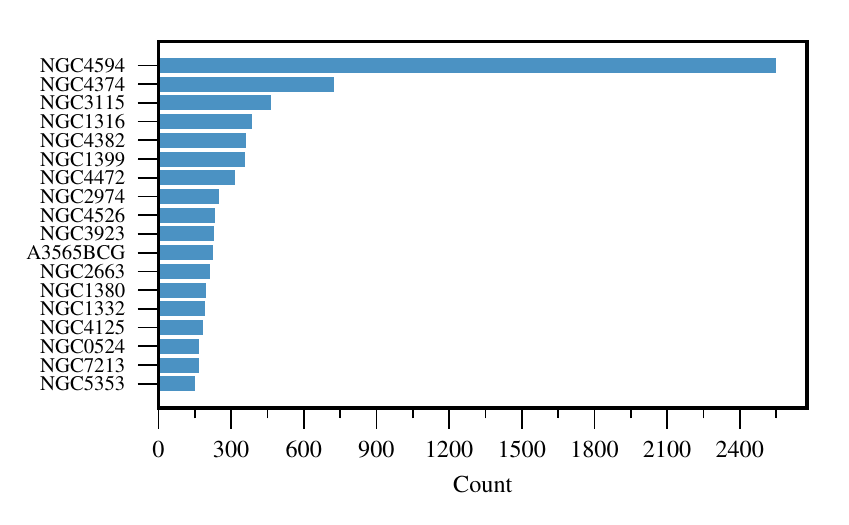}
	\label{fig:hitlist_white}
	}    
   \subfigure[Distance to detected galaxies]
  	{%
	\includegraphics[width=0.45\textwidth]{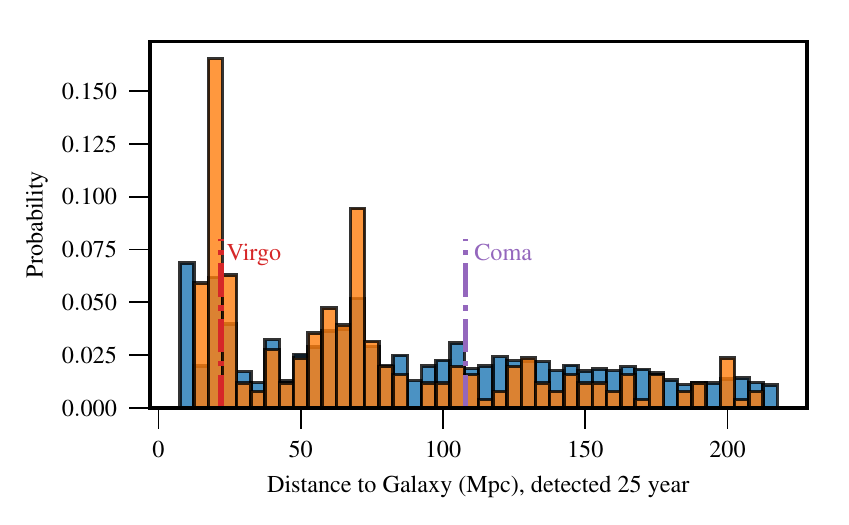}
        }      
    \subfigure[Chirp mass distribution in detected sources]
  	{%
	\includegraphics[width=0.45\textwidth]{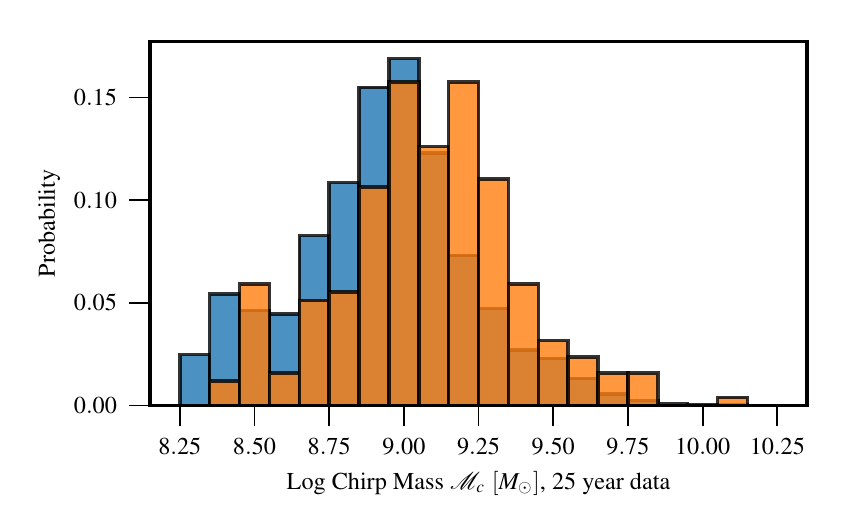}
       \label{fig:chirp}
        }

    \subfigure[Distribution of GW frequencies]
  	{%
	\includegraphics[width=0.45\textwidth]{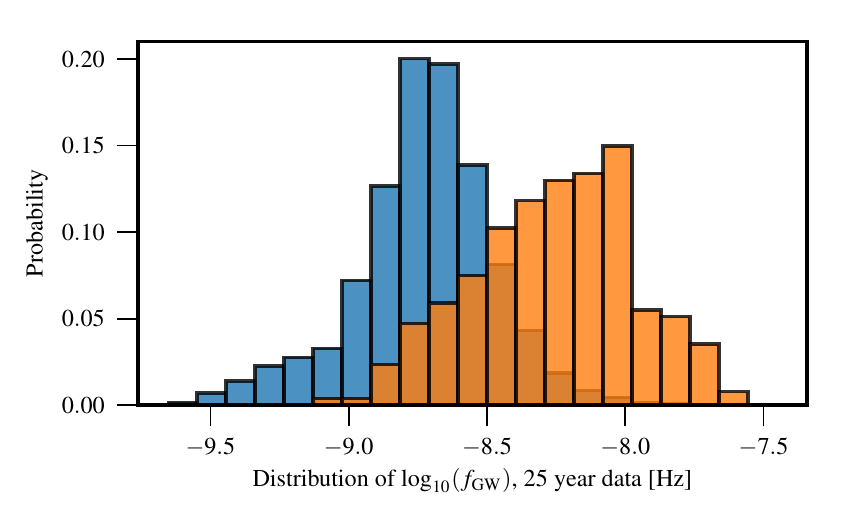}
        }             
   \subfigure[Distribution of strain $h$]
  	{%
	\includegraphics[width=0.45\textwidth]{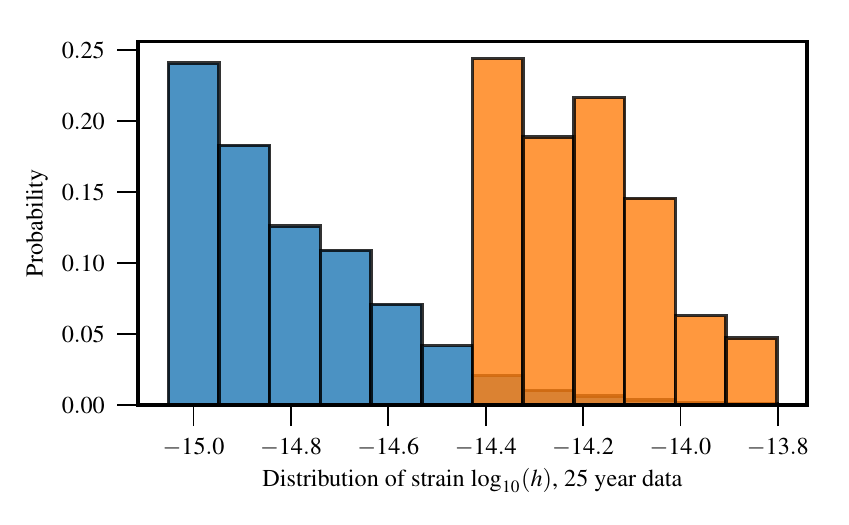}
        }  
  
  \end{center}
\caption{ {\bf Properties of SMBHBs detected with 95\% detection probability and $3\times 10^{-3}$ false alarm probability with a 25 year IPTA dataset}. Blue (orange) histograms are for white (red)-noise pulsars. We find that the selected SMBHB host galaxies using the red-noise projections are more likely to be nearby (in Virgo), at higher GW frequencies, and be very loud.}
\label{fig:allParams}
\end{figure*}

An overview of the detected CGW parameters is given in Figure \ref{fig:allParams}. Interestingly, massive galaxies such as M87 have a lower probability of being selected, since this depends on $t_c \propto \mathcal{M}_c^{-5/3}$. Therefore, binaries in e.g. M104 are more likely to host SMBHBs in the PTA band, Figure \ref{fig:allParams}.

We performed a brief literature search on the likeliest galaxies to host SMBHBs (Supplementary Figures \ref{fig:hitlist_red} and \ref{fig:hitlist_white}) to assess whether they showed signs of merger or a current candidate binary. We find that NGC~3115 is the only object currently under investigation as a candidate binary or recoiling black hole \citep{wrobelnyland}. While many of the other candidates in the top SMBH red-noise candidate list show signs of recent or ongoing merging activity, many galaxies in this mass range are involved in merging, and a more complete comparison between the merger properties of this sample with the general population is beyond the scope of this work.

\section*{Methods}

\textbf{Galaxy Selection from 2MASS}

We select our initial sample from the Two
Mass Redshift Survey (2MRS \citep{HuchraEtAl:2012}).
We first make a very broad selection of objects with $K<-22$ and radial velocity distance $D<250$~Mpc. We then cross-correlate with 
the Crook group catalog \citep{CrookEtAl:2007} to
correct all radial velocities for galaxies in our sample to the radial
velocity of the group. In practice, this has a negligible impact on
the majority of galaxies in the sample, which live at the centers of
their groups. Finally, with group distances and magnitudes in hand, our sample has 5,110 massive early-type galaxies with $K<-25$ mag and $D<225$ Mpc.

To make our sample cleaner, and enable the clean conversion of stellar
mass to an inferred black hole mass, we select only early-type
galaxies (elliptical or S0), using morphologies from Hyperleda \citep{PaturelEtAl:2003}. 
We visually inspected $\sim 1000$ of the galaxies, and found the sample to 
be very clean.

The galaxy catalog is augmented by adding nearby galaxies which host SMBHs with dynamical mass measurements \citep{SchutzMa:2016, MaEtAl:2014}. 33 such galaxies are added, 9 of which were not previously in the 2MASS sample (due to e.g. low $M_K$ luminosity). For these galaxies, we use the measured SMBH mass instead of inferring it from an empirical scaling relation.

\textbf{Supermassive Black Hole Masses} 
We estimate the SMBHB mass by converting the 2MASS K-band luminosity to the stellar mass, take this stellar mass to be the bulge mass, and further apply the $M_\bullet-M_{\mathrm{bulge}}$ scaling relation from Ref \citep{mm13} to obtain the total BH mass. When computing the total SMBHB mass this way, 
we incorporate an intrinsic scatter in the relation, $\epsilon_0$, as follows: each logarithmic realization of $M_\bullet-M_{\mathrm{bulge}}$ is a random draw from a normal distribution with mean $\mu$ from $M_\bullet-M_{\mathrm{bulge}}$, and $\sigma = 0.34$. The answer is exponentiated. Since the exponent is drawn from a normal distribution, it follows that the total BH mass follows a log-normal distribution over many realizations, thus favoring smaller masses. We therefore consider the binary mass to be conservative.

\textbf{IPTA predictions} We add four pulsars per year to the existing 47 IPTA pulsars, chosen from sky locations in the field of view of the current IPTA telescopes. The white noise level is modeled as a combination of radiometer noise and jitter noise.
The radiometer noise is estimated as the harmonic mean of the measured error bars (for each backend and observing frequency) to avoid overestimation due to low signal-to-noise ratio time of arrivals (TOAs), which would have large error bars. Jitter noise is obtained for each observing frequency \citep{lcc+16}. We then compute an infinite frequency TOA uncertainty from the low and high frequency noise estimates in order to simulate dispersion measure fitting. From 2016 onward we add four pulsars per year using the median white noise uncertainty of the existing pulsars in the array, typically around 300 ns. Further, we assume a new wide-band timing backend installed in 2018 at Arecibo and the Green Bank Telescope, which reduces the white noise rms by a factor of $\sim 1.7$.

\begin{figure*}
\begin{center}

	\subfigure[Pessimistic IPTA projections ]{
	\includegraphics[width=0.45\textwidth]{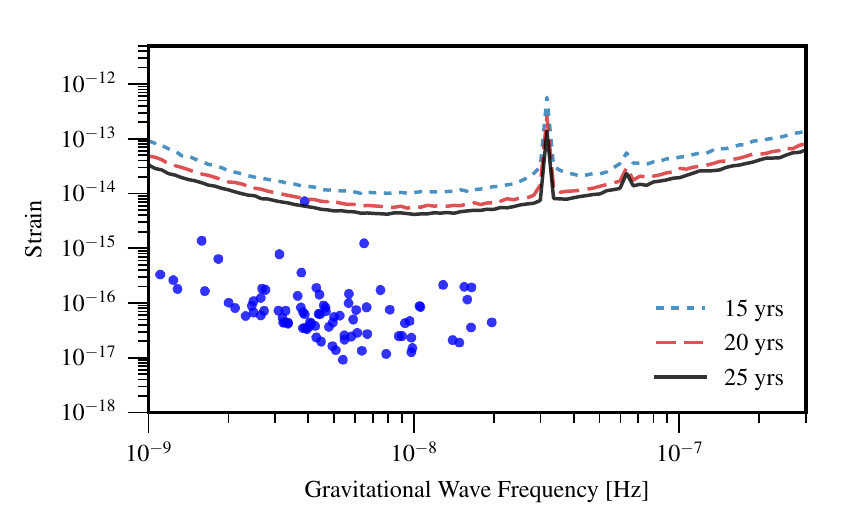}
	\label{fig:pessimisticTTD}
	}
	\subfigure[Optimistic IPTA projections]{
	\includegraphics[width=0.45\textwidth]{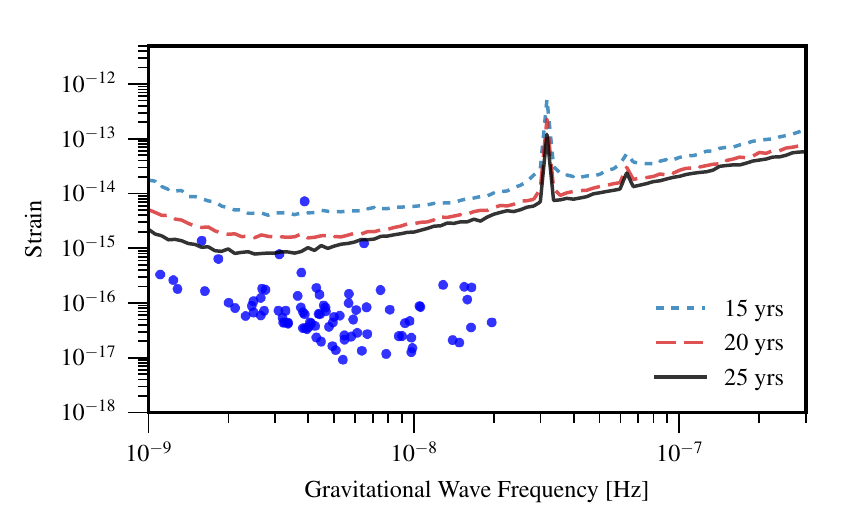}
	\label{fig:optimisticTTD}
	}		
\end{center}
\caption{{\bf Example realization of the local GW sky, with $3\times 10^{-3}$ false alarm probability (FAP) detection curves} Blue dots are GW sources from the same GW sky chosen in Figure \ref{fig:allanis}. This particular realization contains 87 GW sources. Detected sources would lie above the chosen detection curve, for 15, 20, or 25 year datasets. Current PTA datasets are 10-15 years long. (a) Pessimistic outlook where pulsar residuals have a strong red noise component of $4\times 10^{-16}$, emulating the amplitude of an unresolved GWB from models explored in e.g. Ref \citep{SesanaEtAl:2016}. (b) Optimistic outlook with only white noise in the pulsar residuals. Predictions for $5\times 10^{-2}$ and $1\times10^{-4}$ FAPs are in Supplementary Figures \ref{fig:iptaProj_all}.}
\label{fig:iptaProj}
\end{figure*}

This backend upgrade is the dominant factor in the improved white noise detection curves, Figure \ref{fig:optimisticTTD}, due to the fact that the signal-to-noise, $\rho$, of a CGW detection is roughly $\rho \propto \left<NTc/\sigma^2\right>^{1/2}$, assuming the N pulsars have identical intrinsic properties, $T$ is the length of the dataset, $c$ is the cadence of the observation and $\sigma$ is the white noise rms \citep{abb+14}. For red noise with spectral index $\gamma$, $\rho \propto \left<NTc/f^{-\gamma}\right>^{1/2} = \left<NT^{1-\gamma}c\right>^{1/2}$ at $f=1/T$.  There are of course other factors which motivate a large and expanding PTA, including the geometric term from the antenna beam pattern $F^{+,\times}(\hat\Omega) \propto (1+\hat\Omega \cdot \hat p)^{-1} $, where $\hat\Omega$ is the direction of propagation of the GW, $\hat p$ is the direction to the pulsar \citep{det79, abb+14, ms14}, and $+, \times$ is the GW polarization. Therefore, when $\hat\Omega \cdot \hat p \approx -1$ (when the direction to the source, $-\hat\Omega$, is aligned with the pulsar) the response is maximal\footnote{When this alignment is exact there is no GW signal due to surfing effects but when the alignment is almost perfect the response is maximal}, as observed in Figure \ref{fig:allSky}. 

Projections are made for 15, 20 and 25 year datasets with various false alarm probabilities. One can covert from the FAP, $x$, to multiples of the standard deviation $\sigma$ via $x\sigma = $erf$(x/\sqrt(2))$. For example, a FAP of $10^{-4} = 3.9\sigma$, assuming a Gaussian distribution.

\cm{Currently, a full Bayesian analysis is computationally intractable when performing these kinds of detection sensitivity analyses. Since we are performing these analyses as a function of CGW frequency and PTA configuration, we must compute the detection statistic (Bayes factor for Bayesian analysis, FAP for frequentist analysis) millions of times. In comparison to the frequentist FAP statistic, which takes fractions of a second to compute, the Bayes factor computation requires several hours to compute. Thus, for the number of simulations required in this work, a full Bayesian analysis for all simulations is intractable at this time, and we have instead used a frequentist proxy assuming only white noise in order to emulate the possible resolving capability of the Bayesian analysis.}

\begin{figure*}
\begin{center}
      \subfigure[Galaxies per realization hosting PTA SMBHBs]
  	{%
	\includegraphics[width=0.45\textwidth]{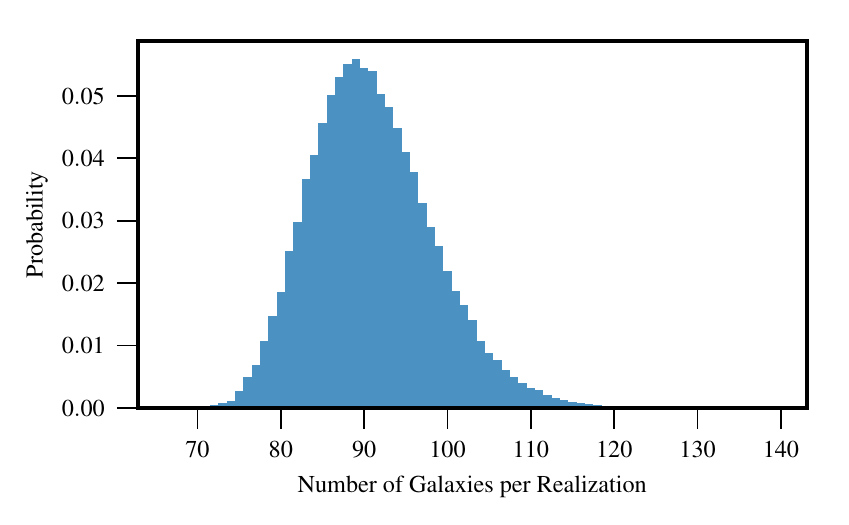}
        }   
       	 \subfigure[Galaxies with stalled SMBHBs]
	{%
	\includegraphics[width=0.45\textwidth]{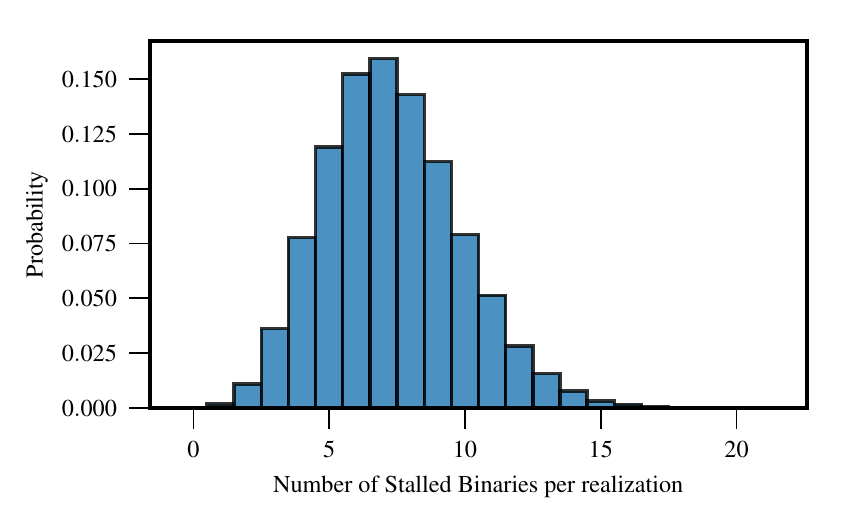}
        }
      
        	\subfigure[Redshift of parent galaxy mergers]{
	\includegraphics[width=0.45\textwidth]{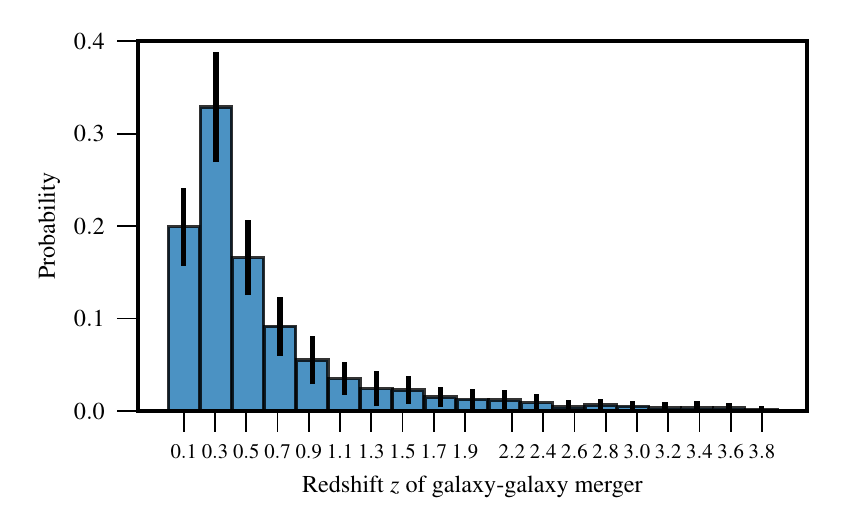}
	\label{fig:binnedZ}
	}
	    
  \end{center}
  \caption{Additional black hole and galaxy information (a) The number of galaxies per realization hosting SMBHBs in the PTA band is $91\pm7$. (b) There are $7\pm2$ stalled SMBHB per realization. (c) Redshift of parent galaxy mergers over all realizations of the 2MASS-based galaxy catalog. Each bin corresponds to the mean value over all realizations, and error bars are 1$\sigma$.} 
  \label{fig:GalStall}
  \end{figure*}

Note that for pulsars with strong red noise, 100 times fewer sources are detected with white-noise dominated pulsars, Figure \ref{fig:iptaProj} and Table \ref{tab:ipta}. 
The presence of unresolved red noise alters the position of the minimum frequency and achievable strain, shifting the lowest GW frequency accessible by PTAs to higher frequencies. Hence, the sources must be much closer (we find that the majority of these are within 20 Mpc) in order to be detected. We also find that, on average, the galaxies in our catalog underwent a major merger at $z=0.3$, Figure \ref{fig:binnedZ}.

\textbf{Generating GW sky maps}
GW sky maps are created by interpolating a set of 128 original data points at 87 GW frequencies. We interpolate between the points using a bivariate spline approximation over a rectangular mesh on a sphere, and project the resulting sky onto a healpix map. This is done for 87 GW frequencies, with the resulting in 87 healpix maps saved as FITS files which are freely available. 
It is possible that more sources could be detected if future data points are extended to a greater range in declination. Interpolation errors close to the poles required us to make a hard cut at declinations of $\pm 70$~degrees, eliminating potential galaxies as CGW sources.

\textbf{Galaxies Hosting SMBHBs in the PTA band} 

\cm{A preliminary study of local potential nanohertz CGW sources was carried out by Ref \citep{spl+14}. The authors assembled a $90\%$ complete galaxy catalog out to $150$~Mpc, including galaxies with SMHBs with $M\geq 10^7~\Msun$. However, in this study it was assumed that there was an equal probability for all galaxies to host a SMBHB, and that this was an equal mass binary}. Moreover, the authors of Ref \citep{RosadoSesana:2014} employed a top-down approach \cm{to predict GW skies} by creating a simulated galaxy catalog, and matching SMBHB merger rates from the Millennium Simulation \citep{SpringelEtAl:2005} to the Sloan Digital Sky Survey \citep{sdss}(SDSS), in an effort to identify the characteristics of potential SMBHB host galaxies. However, SDSS has a limited sky coverage, and does not allow a full-sky investigation of the loudest potential GW sources in the nearby Universe. The work presented here marks the use a galaxy survey to identify local massive galaxies, to assign each galaxy a probability that it hosts a SMBHB based on Illustris galaxy-galaxy merger rates, and to estimate the time to detection of these sources with PTAs.

Here we give an example of how probabilities are assigned to galaxies in the catalog derived from 2MASS, in our bottom-up approach. Consider, for example, galaxy NGC 4594\footnote{NGC 4594 has a dynamically measured SMBH mass, however, for illustration purposes we show how its mass would be assigned via the empirical $M_\bullet-M_\mathrm{bulge}$ relation}, which has $M_K=-25.88$. Through the $M_K-M_*$ empirical scaling relation \citep{Cappellari:2013}, the stellar mass is $M_*=7.03\times 10^{11}~\Msun$, and via $M_\bullet-M_\mathrm{bulge}$ with scatter $\epsilon_0$, a SMBHB with total mass $M =1.92\times 10^9 ~\Msun$. The mass ratio of the binary, $q$, is drawn from a log-normal distribution in [0.25, 1.0], from which we randomly draw $q=0.47$. The chirp mass is therefore $\mathcal{M}_c = 7.69\times 10^8~\Msun$.  

\begin{figure}
\includegraphics[width=.49\textwidth]{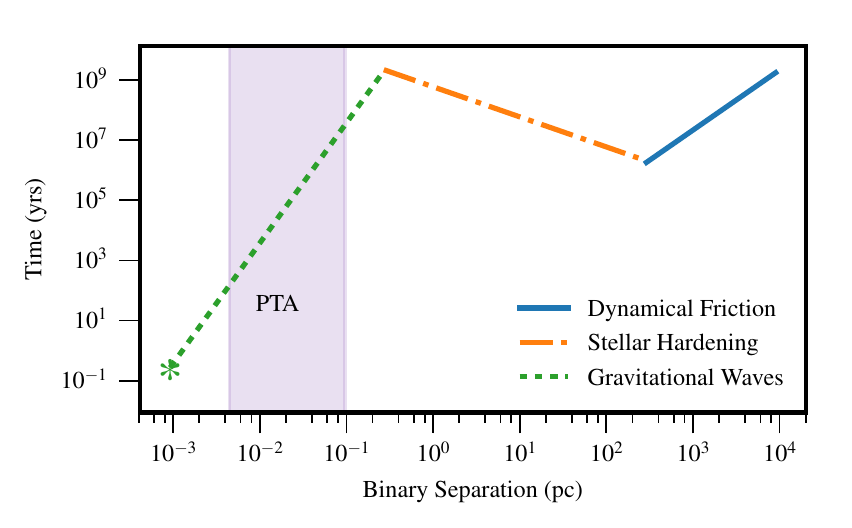}
\caption{{\bf Supermassive black hole binary merger timescales} Here we show the dynamical friction, stellar hardening, and gravitational wave timescales for the worked example with NGC 4594. To obtain the bound of the PTA band, we convert the gravitational wave frequency at the PTA bounds of 1~nHz to 100~nHz to binary separation via  $a=[M/(\pi f_\mathrm{gw})^2]^{1/3}$. We halt the gravitational wave evolution at the binary's innermost stable circular orbit, assuming no spin, $f_\mathrm{ISCO}=1/(\pi 6^{3/2}M)=2.3~\mu$Hz, outside the LISA \citep{LISA:2017} band of $0.1$~mHz to 0.1~Hz.}
\label{fig:BHMerger}
\end{figure}

The dynamical friction timescale \citep{BinneyTremaine} is computed assuming the Coulomb logarithm is $\log(\Lambda)=10$:
\beq
\label{eq:df}
t_\mathrm{df} = 264~\mathrm{Myr}\left(\frac{a}{2~\mathrm{kpc}}\right)^2 \left(\frac{v_c}{250~\mathrm{km/s}}\right)\left(\frac{10^8~\Msun}{M_2}\right)\, ,
\eeq
where $M_2 = qM$ is the mass of the secondary BH, $v_c =\sqrt{2}\sigma$ with $\log \sigma(M_*) = 2.3+0.3\log(M_*/10^{11}\Msun)$ \citep{ZahidEtAl:2016}, and $a$ is the \cm{galaxy's} effective radius, $R_\mathrm{eff}$,  Eq. 4 Ref \citep{DabringhausenEtAl:2008}, Figure \ref{fig:BHMerger}.

\cm{We scale the stellar mass of the galaxy, $M_*$ , by a factor of 0.7, Ref \citep{DeLuciaBlaizot:2007, MFM:2009} (see Supplementary Figure \ref{fig:GalStall}), to estimate the mass of the descendent galaxy when we begin the SMBHB evolution, in the dynamical friction phase. This scaled $M_*$ is also used to estimate the velocity dispersion $\sigma$. }
The parameters drawn are therefore: $M_2=6.14\times10^8~\Msun$, $\sigma=319$~km/s, $R_\mathrm{eff}=7.3$~kpc, which when input in Eq. \eqref{eq:df} yield a dynamical friction timescale of $t_\mathrm{df} = 1.03$~Gyr.

The stellar hardening timescale  \citep{q96} is computed as in Ref \citep{sk15}, Eqs (6) and (7):
\bea
a_{*, \mathrm{gw}} = \left(\frac{64\sigma_{\mathrm{inf}}M_1M_2M }{5 H \rho_{\mathrm{inf}}}\right)^{1/5}  &;&\hspace{0.5cm} t_\mathrm{sh} = \frac{\sigma_\mathrm{inf}}{H\rho_\mathrm{inf} a_{*,\mathrm{gw}}} \, ,
\eea
where $\sigma$ is computed via $M_*-\sigma$  \citep{ZahidEtAl:2016} as above, and $\rho_\mathrm{inf}$ is the density profile evaluated at the influence radius, $r_\mathrm{inf}$ with $\gamma=1$ (corresponding to a Hernquist profile \citep{Dehen:1993}), more details given in Ref \citep{sk15}. Here we assume that the binary eccentricity $e=0$ and the hardening constant $H=15$, and find that the hardening timescale is $t_\mathrm{sh}=2.54$~Gyr.

The sum of the dynamical friction and hardening timescales is $1.03+2.54 = 3.57 $~Gyr, or $z=0.3$ using Planck \citep{Planck2015} cosmological parameters. The cumulative galaxy-galaxy merger rate (Ref \citep{RodriguezGomez+:2015} Table 1) requires only $M_*$ and $z$ as inputs, since the dependence on $\mu_*$ is removed by integrating it between $0.25 \leq \mu_*\leq1$; see Equation \eqref{eq:dNdT}. We scale $M_*$ by 0.7, and find that  is $dN_\mathrm{merg}(M_*,z)/dt = 0.11$.

Finally, we compute the time this binary will spend in the PTA band. The time to coalescence of the binary, $t_c$, is taken from the lower limit of the PTA band: $f_\mathrm{min}=10^{-9}$~Hz, resulting in $t_c = 31.8$~Myr.
By Eq. \eqref{eq:dNdT}, the probability of NGC 4594 hosting a SMBHB in the PTA band in this particular realization is $p = (31.8~\mathrm{Myr})(0.11$~mergers/Gyr) $= 3.5\times10^{-3}$.

\textbf{Computer Code and Data Availability} A series of Jupyter Notebooks which reproduce our figures and all our results are freely available \url{https://github.com/ChiaraMingarelli/nanohertz_GWs}. Through this git repository, we also share the underlying detection curves and GW sensitivity sky maps.

Correspondence and requests for materials should be addressed to Chiara Mingarelli.

\acknowledgments
The authors thank S. Babak, J. Verbiest, D. Kaplan, E. Barr, K. G{\'o}rski and E. Sheldon for useful discussions. The authors are grateful for open source scientific tools for Python \citep{scipy, astropy, numpy, matplotlib, healpy, ipython} and L. Singer's open source plot.py \citep{SingerEtAl:2014}. 
This publication makes use of data products from the Two Micron All Sky Survey, which is a joint project of the University of Massachusetts and the Infrared Processing and Analysis Center/California Institute of Technology, funded by NASA and the NSF.
C.M.F.M. was supported by a Marie Curie International Outgoing Fellowship within the European Union Seventh Framework Programme. S.R.T was partly supported by appointment to the NASA Postdoctoral Program at the Jet Propulsion Laboratory, administered by Oak Ridge Associated Universities and the Universities Space Research Association through a contract with NASA. AS is supported by a URF of the Royal Society. Part of these computations were performed on the Zwicky cluster at Caltech, which is supported by the Sherman Fairchild Foundation and NSF award PHY-0960291. Part of this research was carried out at the Jet Propulsion Laboratory, California Institute of Technology, under a contract with NASA. This work has also been supported by NSF award 1458952. The NANOGrav project receives support from National Science Foundation Physics Frontier Center award number 1430284.

\textbf{Author contributions}
C.M.F.M. modeled the supermassive black hole evolution, developed and ran the Monte Carlo simulations used here to explore their evolution, analyzed the resulting data, produced all of the figures and table, and was the primary author of this paper. C.M.F.M, T.J.W.L and S.B.S. developed the concept of this work.  A.S., C.P.M., S.C. and T.J.W.L. advised on supermassive black hole astrophysics and helped to interpret the results. J.E.G. and S.C. assembled and inspected the galaxy catalog. J.A.E. developed the time to detection methods for the IPTA. S.R.T. helped to develop the methods used to turn continuous gravitational-wave sources into a gravitational wave background, and explore its angular power spectrum.

\section*{Supplementary Materials}

\begin{figure*}
\begin{center}
	\subfigure[Pessimistic IPTA projections, $5\times 10^{-2}$~FAP ]{
	\includegraphics[width=0.45\textwidth]{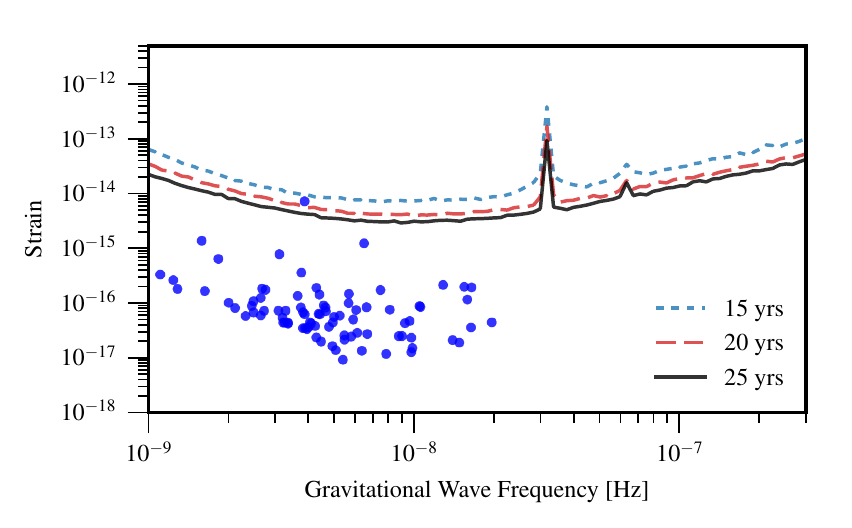}
	\label{fig:pessimisticTTD2sig}
	}
	\subfigure[Optimistic IPTA projections, $5\times 10^{-2}$~FAP]{
	\includegraphics[width=0.45\textwidth]{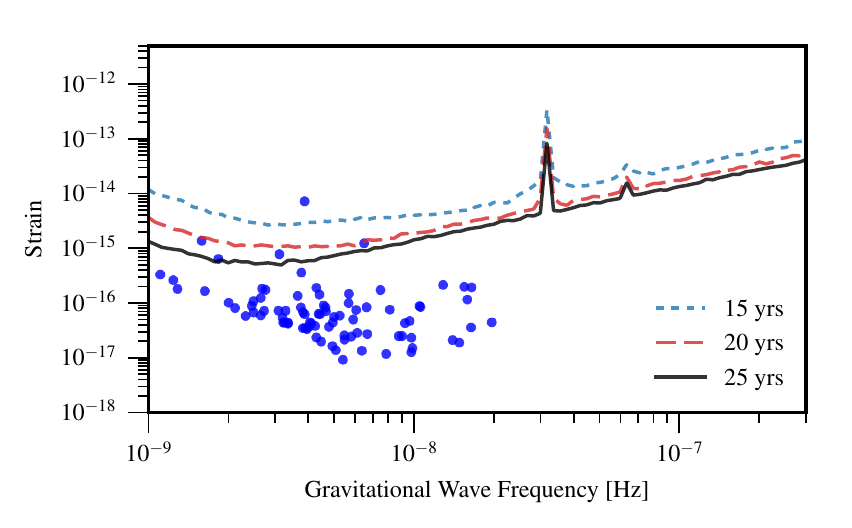}
	\label{fig:optimisticTTD2sig}
	}
	\subfigure[Pessimistic IPTA projections, $1\times 10^{-4}$~FAP ]{
	\includegraphics[width=0.45\textwidth]{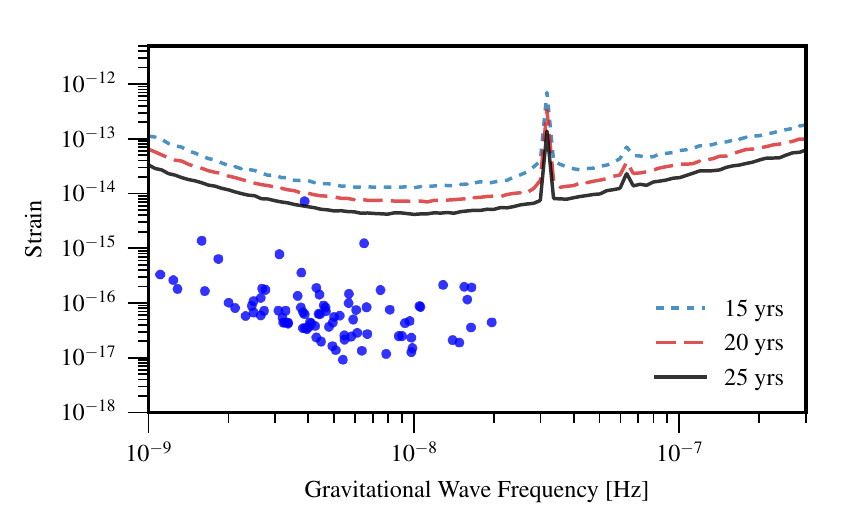}
	\label{fig:pessimisticTTD21e-4}
	}
	\subfigure[Optimistic IPTA projections, $1\times 10^{-4}$~FAP]{
	\includegraphics[width=0.45\textwidth]{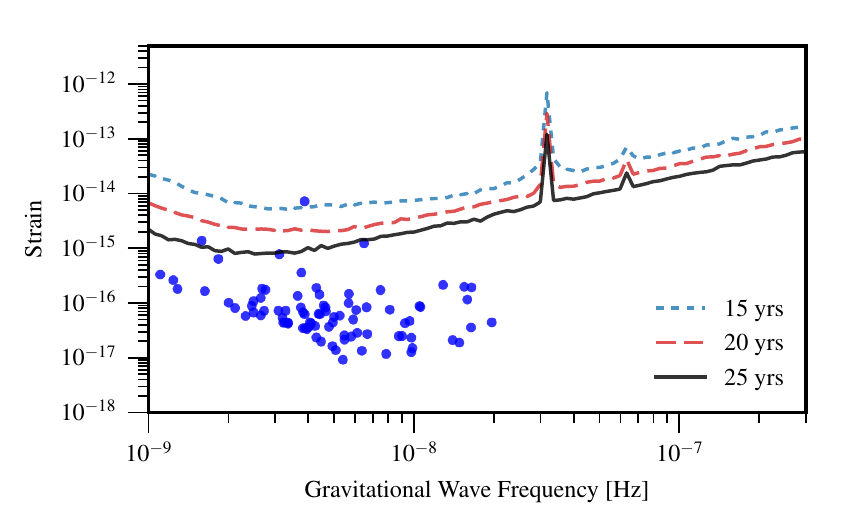}
	\label{fig:optimisticTTD1e-4}
	}	
\end{center}
\caption{{\bf Example realization of the GW sky for IPTA detections with $5\times 10^{-2}$ and  $1\times 10^{-4}$ false alarm probability (FAP)}. The $3\times 10^{-3}$ FAP projections are in Figure \ref{fig:iptaProj}. Blue dots are all the sources from an example GW sky, with detected sources lying above the curve. The pessimistic outlook assumes pulsar residuals have a strong red noise component of $4\times 10^{-16}$, equivalent to an unresolved GWB, whereas the optimistic outlook assumes only white noise in the pulsar residuals.}
\label{fig:iptaProj_all}
\end{figure*}

\bibliographystyle{apj}
\bibliography{bib, apjjabb}

\end{document}